\documentclass[twocolumn,showpacs,prb,10pt]{revtex4}
\usepackage[english]{babel}
\usepackage{amsmath,amssymb}
\usepackage{times}
\usepackage{epsfig}

\begin{document}

\title {Anomalous scaling of conductivity in integrable fermion
systems} 
\author{P. Prelov\v sek$^{1,2}$, S. El Shawish$^{1}$,
X. Zotos$^3$, and M. Long$^4$} 
\affiliation{$^1$J.\ Stefan Institute,
SI-1000 Ljubljana, Slovenia} 
\affiliation{$^2$ Faculty of Mathematics
and Physics, University of Ljubljana, SI-1000 Ljubljana, Slovenia}
\affiliation{$^3$ Department of Physics, University of Crete and
Foundation for Research and Technology-Hellas, P.O. Box 2208, 71003
Heraklion, Greece} 
\affiliation{$^4$ Department of Physics, University
of Birmingham, Edgbaston, Birmingham B15 2TT, United Kingdom}
\date{\today}

\begin{abstract}
We analyze the high-temperature conductivity in one-dimensional
integrable models of interacting fermions: the $t$-$V$ model
(anisotropic Heisenberg spin chain) and the Hubbard model, at
half-filling in the regime corresponding to insulating ground
state. A microcanonical Lanczos method study for finite size systems 
reveals anomalously large finite-size effects at low frequencies 
while a frequency-moment analysis indicates a finite d.c. conductivity. 
This phenomenon also appears in a prototype integrable 
quantum system of impenetrable particles, representing
a strong-coupling limit of both models.
In the thermodynamic limit, the two results could converge to a 
finite d.c. conductivity rather than an ideal conductor or insulator 
scenario.

\end{abstract}

\maketitle

\section{Introduction} 

Transport of strongly interacting fermions in one-dimensional (1D)
systems have been so far the subject of numerous theoretical as well
as some experimental studies \cite{zprev}. While the ground-state and
low-temperature properties, following the Luttinger-liquid
universality, are well understood, the transport properties still lack
some fundamental understanding regarding the role of fermion
correlations.  It has become evident in recent years, that with
respect to transport (in contrast to static quantities) integrable
many-fermion models behave very differently from nonintegrable ones
\cite{cast,zprev}. Some basic 1D fermion models are integrable, as the
$t$-$V$ model (equivalent to the anisotropic Heisenberg spin model)
and the Hubbard model, and reveal in the metallic regime
dissipationless transport at finite temperature $T>0$, well founded
due to the relation to conserved quantities \cite{zot1,zot2,fabr}.  The
transport in the 'insulating' regime of integrable models, however,
has been controvertial and is the issue of this paper.

Let us concentrate on the dynamical conductivity in 1D system
\begin{eqnarray}
\sigma(\omega)&=& 2 \pi D \delta(\omega) + \sigma_{reg}(\omega),
\nonumber \\
\sigma_{reg}(\omega>0)&=& \frac{1-e^{-\beta \omega}}{\omega L}
{\mathrm Re}\int_0^\infty dt e^{i\omega t} \langle j(t) j(0) \rangle~,
\label{sigma}
\end{eqnarray}
where $j$ is the (total) particle current operator, $\beta=1/T$ and
$L$ is the number of sites in the chain (we set everywhere
$k_B=\hbar=e_0=1$ as well as lattice spacing $a_0=1$). At finite $T$
the charge stiffness (referred to also as Drude weight) $D(T)$
measures the dissipationless component in the response, while
$\sigma_{reg}(\omega)$ is the 'regular' part. The requirement that the
ground state is insulating \cite{kohn} is $D_0=D(T=0)=0$. In the
insulating regime there are still several alternative scenarios for
the behavior at finite temperatures.  The system can at $T>0$ behave
as: a) an 'ideal conductor' with $D(T)>0$, b) a 'normal resistor' with
$D(T)=0$ but $\sigma_0= \sigma(\omega \to 0)>0$, and c) an 'ideal
insulator' with $D(T)=0$ and $\sigma_0=0$.

A well-known $T=0$ insulator is the $t$-$V$ model at half filling and
$V/t>2$. The model is equivalent in this regime to an easy-axis
anisotropic XXZ Heisenberg model with $\Delta >1$.  It has been shown
by one of the present authors \cite{zoto} that $D(T>0)$ is finite for
$V/t<2$ but decreasing towards $D(T>0)=0$ at $V/t=2$. This gives a
strong indication that $D(T)=0$ in the whole regime $V/t>2$, 
although there are also alternative interpretations \cite{pscc,bren}. 
The present authors speculated in this case on a possible realisation
of an 'ideal insulator' \cite{zot1} where also $\sigma_0(T>0)=0$. The
argument is based on the observation that at least in the $V/t \to
\infty$ limit the soliton-antisoliton mapping can be applied, where
the eigenstates cannot carry any current. However, the issue proved to
be more involved. Note that the transport of gapped spin systems
described by the quantum nonlinear sigma model, when treated by a
semiclassical approach \cite{sach} (mapping to a model of classical
impenetrable particles) indicates a 'normal conductor' with a finite
diffusion constant and $\sigma_0(T>0)>0$. On the other hand, a Bethe
ansatz approach \cite{fuj} concludes to a finite Drude weight
$D(T>0)>0$ and thus ballistic transport.  It should be reminded that
the $V/t=2$ case, corresponding to the most studied isotropic
Heisenberg model, is marginal situation, with the long-standing open
question whether the diffusion constant (studied mostly at $T \to
\infty$) in this model is finite \cite{carb,fabr}.  Another 
prominent $T=0$ insulator is the Hubbard model at half filling. Here
even the question of $D(T>0)$ is controvertial.  On the basis of Bethe
ansatz results \cite{fuji} and Quantum Monte Carlo simulations
\cite{kirc} it is claimed that $D(T>0)>0$, i.e. an 'ideal conductor'
situation. More recent analytical considerations \cite{pere} seem to favor
$D(T)=0$.

The aim of this paper is to present numerical evidence that the
dynamical conductivity $\sigma(\omega)$ in the insulating regime of
several integrable 1D models is indeed very anomalous. We consider in
this context three 1D models: the $t$-$V$ model, the Hubbard model and
a related model of impenetrable particles. First, finite-size scaling
of results for all mentioned models indicates that indeed $D(T>0)=0$
(whereby the evidence is somewhat less conclusive for the Hubbard
model).  Moreover, we show that on the one hand small-system results
reveal large pseudogap features in $\sigma(\omega \sim 0)$ and large
finite size effects extending to high frequencies; on the other hand,
after the finite-size scaling in the thermodynamic limit is performed, 
the results could be consistent with a 'normal' and featureless
$\sigma(\omega)$ found by a frequency-moment analysis. In this respect
the behavior is very different from the one in nonintegrable quantum
many-body models where even in small-size systems a 'normal' diffusive
behavior is very evident \cite{zot1,zot}.

The paper is organized as follows. In Sec.~II we present two
alternative numerical methods used to analyse the dynamical
conductivity $\sigma(\omega)$: the microcanonical Lanczos method and
the method of frequency moments. In Sec.~III results for three different
1D models in the insulating regime are presented and discussed: the
$t$-$V$ model at half-filling, the Hubbard model at half-filling and
the model of impenetrable particles.

\section{Numerical methods}

Microscopic models considered in this paper are 1D tight-binding
models with the hopping only among nearest neighbors.  We investigate
within these models the dynamical charge conductivity $\sigma(\omega)$
(in the case of impenetrable particles the related spin conductivity
$\sigma_s(\omega)$) at $T \to \infty$ with an emphasis on the low
$\omega \to 0$ behavior. The first approach we apply is the full exact
diagonalization (ED) of the Hamiltonian on a lattice with $L$ sites and 
periodic boundary conditions (p.b.c.) taking into account the number
of fermions $N$ and the wavevector $q$ as good quantum numbers. E.g.,
this allows for an exact solution of $\sigma(\omega)$ up to $L=20$ for
the $t$-$V$ model. Larger systems can be studied using the Lanczos
method of ED.

Particularly appropriate at large enough $T$ is the microcanonical
Lanczos method (MCLM) \cite{long}. The MCLM uses the idea that
dynamical autocorrelations (in a large enough system) can be evaluated
with respect to a single wavefunction $|\Psi\rangle$ provided that the
energy deviation
\begin{equation}
\delta \epsilon = (\langle\Psi|(H-\lambda)^2|\Psi\rangle)^{1/2}
\end{equation}
is small enough. Clearly, $\lambda$ determines here the temperature
$T$ for which $|\Psi\rangle$ is a relevant representative. Such
$|\Psi\rangle$ can be generated via a first Lanczos procedure using instead
of $H$ a modified projection operator $P=(H-\lambda)^2$, performing $M_1$
Lanczos steps to get the ground state of $P$. The dynamical
correlations are then calculated using the standard Lanczos procedure
for dynamical autocorrelation functions, where the modified $|\tilde
\Psi\rangle = j|\Psi\rangle$ is the starting wavefunction for the 
second Lanczos iteration with $M_2$ steps generating the continued
fraction representation of $\sigma(\omega)$.  The main advantage of
the MCLM is that it can reach systems equivalent in size to the usual
ground-state calculations using the Lanczos method. For details we
refer to Ref.[17]; e.g., the largest available size for the
$t$-$V$ model is thus $L=28$.

Besides the $\sigma(\omega)$ spectra it is instructive to also show
the normalized integrated intensity $I(\omega)$. In tight-binding
models with n.n. hopping the (optical) sum rule for $\sigma(\omega)$
is given by $\langle -T \rangle/2L$ where $T$ is the kinetic term in
the Hamiltonian.  Hence $I(\omega)$ can be expressed as
\begin{equation}
I(\omega)=D^* + \frac{2L}{\langle - T \rangle} \int_{0^+}^{\omega} 
d\omega' \sigma(\omega'), \label{eq2}
\end{equation} 
which is monotonously increasing function with the limiting value
$I(\omega \to \infty)=1$ and well defined even for small
systems. Here, $D^*=2LD/\langle -T \rangle$. It should be noted that
in a full ED calculation the Drude part $D$ appears strictly at
$\omega=0$, Eq.~(\ref{sigma}), while in the MCLM it spreads into a
window $\delta\epsilon \ll t$ governed by the number of Lanczos steps
$M_1$. Still, choosing large enough $M_1 \sim 1000$, $\delta\epsilon$
becomes very small, hence we get well resolvable Drude
contribution. Typically we use in the calculations presented here $M_1
= 1000$, $M_2 = 5000$. In order to get smooth spectra especially for
small system sizes, we additionally performed an averaging over
$N_{\lambda}$ different $\lambda$ with respect to the normal Gaussian
distribution. Typically, we used $N_{\lambda}\sim 20$ for smallest $L$
and $N_{\lambda}\sim 1$ for largest $L$ presented in figures below.

As will be evident from results furtheron $\sigma(\omega)$ exhibits
huge finite-size effects. The latter are clearly a consequence of the
integrability since nonintegrable models do not exhibit such
features. In order to avoid such finite-size phenomena we also perform 
an alternative analysis using the method of frequency moments (FM).
It is well known that at $T=\infty$ one can calculate for
$\sigma(\omega)$ exact frequency moments $m_{2k}=\pi \mu_{2k}/T$ as
\begin{equation}
\mu_{2k}={\rm Tr}([H,[H,\cdots, [H,j]]\cdots ]j)/{\rm
Tr}(1). \label{eq3}
\end{equation}
Moments correspond here to an infinite system $L \to \infty$ and could
be evaluated at fixed fermion concentration $n=N/L$ using the linked
cluster expansion and the diagrammatic representation
\cite{ohat,mori}.  Only clusters containing up to $k+1$ particles can
contribute to $\mu_{2k}$ in an infinite system for models with
n.n. connections only.  However, an analytic calculation of moments
for larger $k$ becomes very tedious. Hence we use the fact that {\it
exact moments for an infinite system} can be obtained also via the ED
results for small-system (with p.b.c.) \cite{sur} provided that the
system size $L$ is large enough. I.e.,
\begin{equation}
\mu_{2k}= \frac{1}{\Omega}\sum_{N=0}^{L} \sum_{m,l}f^N 
(\epsilon_{Nm}-\epsilon_{Nl})^{2k}
|\langle \Psi_{Nm}|j|\Psi_{Nl}\rangle |^2, \label{eq4}
\end{equation}
where $|\Psi_{Nm}\rangle$ refer to eigenstates for $N$ fermions. In
Eq.~(\ref{eq4}) $\Omega =\sum_N N_{st}(N) f^N$ and fugacity $f={\rm
exp}(\mu/T)$ can be related to the density
\begin{equation}
n=\frac{1}{\Omega L}\sum_N N N_{st}(N) f^N,
\end{equation}
where $N_{st}$ is the number of states for given $N$. Let us
illustrate the feasibility of FM method for the 1D $t$-$V$ model.
Performing full ED for all fillings $0<N<L$ on a ring we get exactly
FM up to $k=L/2-1$ whereby even higher $k>L/2-1$ moments are quite
accurate. Using full ED for $L=20$ we thus reach for the $t$-$V$ model
exactly up to $\mu_{18}$.

Next step is to reconstruct spectra $\sigma(\omega)$ from $\mu_{2k}$
with $k=0,K$. There are various strategies how to get the spectra most
representative for $K \to \infty$, expecting a smooth function
$\sigma(\omega)$. We follow here the procedure proposed by Nickel
\cite{nick}. First, a nonlinear transformation $\omega=z+\zeta^2/z$ is
performed where $\zeta$ is chosen as the largest eigenvalue in a
truncated continued fraction representation of $\sigma(\omega)$.  For
$\sigma(\omega)$ then a Pad\' e approximant $[K_1/K_2]$ is found in
terms of functions of the novel variable $z$.

It should, however, be noted that FM are less sensitive to the
low-$\omega$ regime, so a possible price to pay is an uncertainty in
the low frequency results. In this respect MCLM and FM results really
yield an alternative view of low-$\omega$ dynamics.

\section{Results}

\subsection{$t$-$V$ model}
Let us first analyse the 1D $t$-$V$ model for interacting spinless
fermions,
\begin{equation}
H=-t\sum_i (c^{\dagger}_{i+1} c_{i} + \mathrm{h.c.}) + V\sum_i n_i
n_{i+1}, \label{eq1}
\end{equation}
with the repulsion $V$ between fermions on n.n. sites and the
corresponding current operator 
\begin{equation}
j=-t\sum_i (ic^{\dagger}_{i+1} c_{i} + \mathrm{h.c.}). 
\end{equation}
At half-filling, i.e., at the fermion density $n=1/2$, the ground
state is metallic for $V/t<2$ and insulating for $V/t>2$. Note that by
introducing a fictitious magnetic flux via the substitution $t \to t
e^{i\phi}$ the model turns into the anisotropic XXZ Heisenberg model
for $\phi=\pi/L$ and even number of fermions.

In the following we present only results in the limit $T \to
\infty$. From Eq.~(\ref{sigma}) it follows that $\sigma$ 
scales in this limit as $\beta$, hence we present in Fig.~1a
$\sigma(\omega)/\beta$, calculated for even number of fermions for
$V/t=4$ and various sizes $L=16-28$. Results for $D/\beta$ are plotted
in the inset of Fig.~1b and reveal an exponential decrease with $L$,
which is at the same time a challenging test for the feasibility and
the sensitivity of the MCLM at larger sizes $L$.

From results in Figs.~1a,b several observations follow: a) the
dissipationless component $D$ becomes negligible at large $L$ and the 
extrapolated value for $L =\infty$ is consistent with $D=0$~ \cite{zoto}, b)
there is a pseudogap at low $\omega$ followed by a pronounced peak at
$\omega=\omega_p$ and damped oscillatory features at
$\omega>\omega_p$, almost up to the bandwidth $\sim 4t$, c) the peak
and accompanied oscillations move downward with the system size
approximately as $\omega_p \propto 1/L$, d) the pseudogap in
$\sigma(\omega \to 0)$ is compensated by the peak intensity as evident
from the integrated $I(\omega)$ which is essentially independent of
$L$ for $\omega > \omega_p$, e) $I(\omega<t)$ could approach
$I(\omega)\sim \sigma_0 \omega$ for large $L$, indicative of a
'normal' d.c. conductivity $\sigma_0$ in the thermodynamic limit.

\begin{figure}[htb]
\centering
\epsfig{file=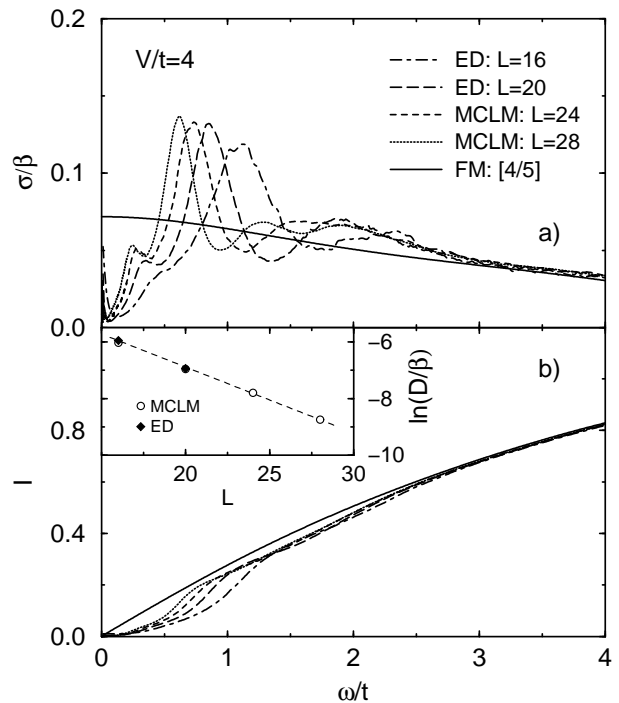,width=80mm,angle=0}
\caption{a) Conductivity $\sigma(\omega)/\beta$ 
and b) integrated normalized $I(\omega)$ at $T \to \infty$ within the
1D $t$-$V$ model with $V/t=4$, obtained using the ED and the MCLM for
systems with length $L$, and the frequency moment expansion. The inset
shows $\mathrm{ln} (D/\beta)$ vs. $L$, where the line is guide to the
eye.  }
\label{fig1}
\end{figure}
 
When applying the FM method to the $t$-$V$ model we get
\begin{equation}
\Omega=(1-n)^{-L}, \qquad f=n/(1-n).
\end{equation}
Using full ED for $L=20$ we reach exactly up to $\mu_{18}$.  In
Figs.~1a.b we diplay results for $\sigma(\omega)$ obtained via the FM
using $K=9$ and the corresponding [4/5] Pad\' e approximant. The FM
method proves to be very stable in particular with respect to the most
interesting and sensitive value $\sigma_0$. Namely the latter varies
only slightly between, e.g., [3/3] and [4/5] Pad\' e approximant.
Results confirm the overall agreement of MCLM and FM-method spectra
apart from evident finite-size phenomena at $\omega<\omega_p$. It
should be, however, mentioned that there are still some nonessential
differences between $I(\omega)$ results even at higher
$\omega>\omega_p$ since the MCLM results are for fixed fermion number
$N=L/2$ whereas the FM corresponds to a grandcanonical averaging over
all $N$ so that even lowest moments differ slightly. The general
conclusion of the FM approach is that it does not show any sign of
pseudogap features and thus favors quite featureless $\sigma(\omega)$
with finite $\sigma_0$. Essentially the same results are reproduced
for $\sigma(\omega)$ analysing FM using the maximum-entropy method
\cite{mead}.

\subsection{Hubbard model}

Next let us consider the 1D Hubbard model
\begin{eqnarray}
H&=&-t \sum_{i,s} (e^{i\phi} c_{i+1,s}^{\dagger} c_{is} +
\mathrm{h.c.})  + U \sum_i n_{i\uparrow} n_{i\downarrow}, \nonumber\\
&&j=-t \sum_{i,s} (i e^{i\phi} c_{i+1,s}^{\dagger} c_{is} +
\mathrm{h.c.}),
\label{eq5}
\end{eqnarray}
where we take into account a possible fictitious flux $\phi$.  We
study the model at half filling $n=N/L=1$ where the ground state is
insulating, i.e. $D_0=0$, for all $U>0$. In the limit $L\to \infty$
the behavior should not depend on $\phi$. Nevertheless in small
systems low-$\omega$ features, in particular $D(\phi)$, depend on
$\phi$. We present here calculations within the Hubbard model using
the ED and the MCLM at $\phi=\pi/(2L)$ since in this case $D(\phi,T)$
is at maximum. Relative to the $t$-$V$ model, smaller sizes are
reachable for the Hubbard model at $n=1$, i.e., we investigate $L=10$
performing full ED, while with the MCLM systems up to $L=16$ can be
studied.

\begin{figure}[htb]
\centering
\epsfig{file=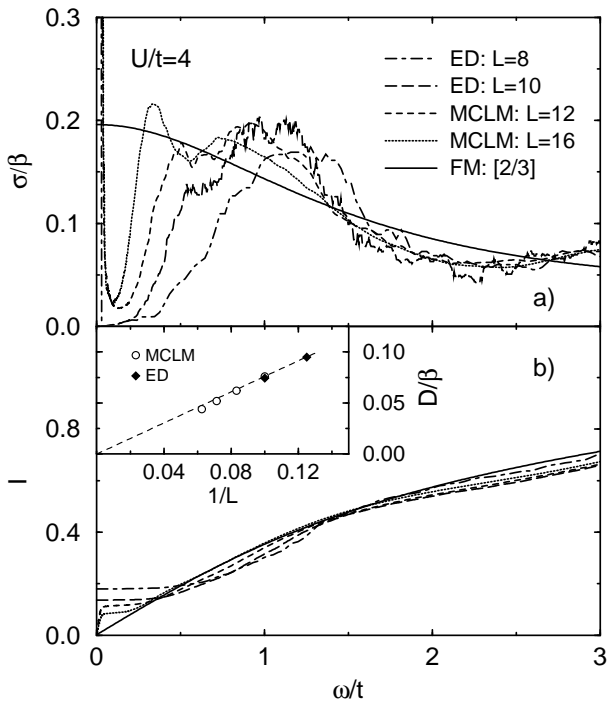,width=80mm,angle=0}
\caption{a) $\sigma(\omega)/\beta$ and b) 
$I(\omega)$ within the 1D Hubbard model with $U/t=4$, obtained via the
ED, the MCLM (finite $L$) and the FM method. The inset shows $D/\beta$
scaled vs. $1/L$, whereby the line is guide to the eye.}
\label{fig2}
\end{figure}

Results for the intermediate case $U/t=4$ and again $T \to\infty$ are
shown in Figs.~2a,b. We note that several features are similar to
results for the $t$-$V$ model: a) $D$ decreases with $L$, b) a
pseudogap appears for $\omega<\omega_p$, c) large finite size effects
extend up to frequencies of the order of the bandwidth, d) the
pseudogap scale appears to close with the increasing system size.

However, the dependence of $D(L)$ is not exponential, but the scaling
appears to follow $D \propto 1/L$ (see the inset of Fig.~2b). Although
with less certainty than within the $t$-$V$ model we could again
support the limiting value $D(T)=0$. Also, $I(\omega)$ tends with
increasing $L$ to $I \sim \sigma_0 \omega$ for $\omega<t$, here
approaching from higher values in contrast to Fig.~1b. In spite of
differences to the $t$-$V$ model, results scaled to the thermodynamic
limit could be consistent with a smooth $\sigma(\omega)$ and a finite
$\sigma_0$.

We also perform the FM analysis, using exact ED results for systems
with up to $L=10$ and $0<N<2L$. Here, we use 
\begin{equation}
\Omega=(1+f)^{2L}, \qquad f=n/(2-n). 
\end{equation}
The analysis is accurate up to $\mu_{10}$ and corresponding [3/2]
Pad\' e approximants. This is barely enough to reproduce gross
features of limiting $\sigma(\omega)$, nevertheless results are in
agreement with previous conclusions for the $t$-$V$ model.

\subsection{Impenetrable particles}

The above results indicate that integrable models in the 'insulating'
regime share similar features in the dynamical conductivity $\sigma(\omega)$. 
It has already been proposed \cite{zot1} that it is helpful to consider the
large interaction limit, i.e., $V \gg t$ and $U \gg t$, where the
dynamics of both models is simplified but remains highly
nontrivial. For a half-filled band in this limit we are dealing with an 
excitation spectrum composed of 
split subspaces with a fixed number $N_s$ of oppositely
charged ``soliton-antisoliton" ($s \bar s$) pairs. 
In such a limit, the solitons/antisolitons - doubly occupied/empty sites 
in the 
Hubbard model, occupied/empty n.n. sites in the $t$-$V$ model - behave as
impenetrable quantum particles, since their crossing would require
virtual processes with $\Delta E=U,V$.  

The simplest prototype model which incorporates the same physics - that 
of a system with two species of impenetrable particles - is the 1D $t$-model,
\begin{equation}
H=-t\sum_{i s}(\tilde{c}^\dagger_{i+1,s}\tilde{c}_{is}+
\mathrm{h.c.}), \label{eq6}
\end{equation}
where projected fermion operators take into account that the double
occupation of sites is forbidden; the two species of particles are 
represented by the up/down spin fermions.
Thus we consider within the $t$-model the spin current
\begin{equation}
j_s=t\sum_{i s}(i s\tilde{c}^\dagger_{i+1,s}\tilde{c}_{is}+ 
\mathrm{h.c.}),
\end{equation}
and the corresponding spin diffusivity $\sigma_s(\omega)$. 

The only relevant parameter within the $t$-model is the electron
density $n=n_\uparrow+n_\downarrow$, where $0<n<1$ and of interest is
the paramagnetic case $n_\uparrow=n_\downarrow$.  The model
(\ref{eq6}) is also exactly solvable. Moreover, the electron current
$j$ commutes with $H$, while the spin current $j_s$ does not. It is
plausible that in an unpolarized ring, $N_\uparrow=N_\downarrow$,
exact eigenstates do not carry any spin current, i.e., $\langle
\Psi_n|j_s|\Psi_n \rangle=0 $, and hence $D(T) \equiv 0$.  This
becomes clear by introducing the fictitious flux by $t \to t
e^{i\phi}$.  Particles cannot cross, so all eigenergies
$\epsilon_n$ are independent of $\phi$. Since $D(T)$ can be related 
\cite{cast} to $\partial^2 \epsilon_n/
\partial \phi^2 $ this leads to $D(T) \equiv 0$.  Still,
this does not preclude $\sigma_s(\omega>0)>0$, since $\langle
\Psi_n|j_s|\Psi_m \rangle \neq 0$ in general.

We study $\sigma_s(\omega)$ within the $t$-model again using the same
methods. With the full ED we reach $L=12$ while with the MCLM up to
$L=20$ sites. For the presentation we choose the quarter-filled case,
$n=1/2$, where most systems are available, $L=8,12,16,20$. Results are
shown in Figs.~3a,b. As expected, finite-size features are very
similar to those in Figs.~1,2, apart from $D \equiv 0$. The pseudogap
is pronounced even more clearly, with the peak frequency $\omega_p
\propto 1/L$. Particularly powerful for this model is the FM
method. Namely, from the full ED results we can evaluate exactly
moments up to $\mu_{20}$. Since there is a single characteristic scale
$t$, the structure of $\sigma_s(\omega)$ is simpler and better
reproducible via the FM method.  Results corresponding to [5/5] Pad\'
e approximant are presented in Figs.~3a,b and again indicate on a
'normal' diffusivity in the thermodynamic limit.

\begin{figure}[htb]
\centering
\epsfig{file=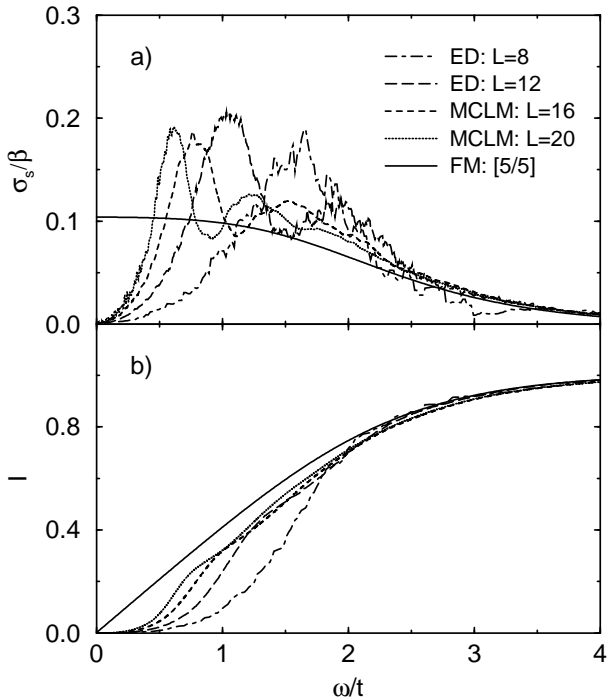,width=80mm,angle=0}
\caption{ a) spin diffusivity $\sigma_s(\omega)/\beta$ and b) the integrated
$I(\omega)$ within the 1D $t$-model at $n=1/2$, obtained via the ED,
the MCLM and the FM method.}
\label{fig3}
\end{figure}

\section{Conclusions}

Let us summarize and comment obtained results. We have shown that all
considered 1D integrable models of interacting fermions share several
common features:

\noindent  a) The charge stiffness is either $D \equiv 0$
($t$-model) or appears to scale to zero, whereby the evidence is
stronger for the $t$-$V$ model. Whereas the vanishing $D$ is easy to
understand for impenetrable particles, it is a highly nontrivial
statement for the other two models
\cite{pscc,bren,fuji,kirc,pere}. The observations can be rationalized
in a way that the $t$-$V$ and Hubbard model at half-filling in the
thermodynamic limit $L \to \infty$ remain to behave as in the limit
$V,U \to \infty$ where solitons and antisolitons cannot cross.

\noindent b) The pseudogap is pronounced for
finite-size systems whereby the finite-size peak scales as $\omega_p
\propto 1/L$. 

\noindent c) The extrapolation to the thermodynamic limit could be
compatible with a rather featureless and regular $\sigma(\omega \sim
0)$ and thus finite $\sigma_0$.  With respect to the last two points
the ED (including MCLM) and FM methods are complementary. Whereas the
FM method (valid for an $L\to \infty$ system) cannot detect
finite-size effects and appears to converge to a featureless
$\sigma(\omega)$, the ED methods are evidently sensitive to the effect
of p.b.c.  at finite $L$.

A fundamental question raised by these observations is, whether the
large finite size effects observed at low frequencies are reflected to
the dynamics of bulk systems and in particular, which features of the
conductivity (e.g.  $q$,$\omega$ -dependence) might be singular.

We restricted our results to $T\rightarrow \infty$. 
The latter is clearly most
convenient for the FM method. Nevertheless, from the MCLM results
considered at finite but high $T$ there appears no evidence for any
qualitative change on behavior.  We also presented results for a single
parameter for the $t$-$V$ and Hubbard model, and one filling for the
$t$ model, respectively.  One cannot expect any essential difference
for other values within the 'insulating' regime, although numerical
evidence becomes poorer, e.g., on approaching $V \to 2~t$ within the
$t$-$V$ model. Clearly, the most challenging case is $V=2~t$,
corresponding to the isotropic Heisenberg model. Our results reveal an
increase of $\sigma_0$ on approaching $V=2~t$. Still we are not able
to settle the well-known dilemma \cite{carb,fabr} whether $\sigma_0$
remains finite or diverges in this marginal case.

We thank N. Papanicolaou for helpful discussions. Authors (P. P.,
S. El S.) acknowledge the support of the Slovene Ministry of
Education, Science and Sports, under grant P1-0044.

\end{document}